\documentclass[aps,prb,superscriptaddress,twocolumn,showpacs,amsmath,amssymb]{revtex4}
\usepackage{graphicx}
\usepackage{epsfig}
\usepackage{pslatex}
\usepackage{natbib}

\newcommand{\chem}[1]{\ensuremath{\mathrm{#1}}}

\begin{document}

\title{Muon spin relaxation investigation of magnetic ordering in the hybrid organic-inorganic perovskites \chem{[(CH_{3})_{2}NH_{2}]}{\it M}\chem{(HCOO)_{3}}, $M=$~\chem{Ni, Co, Mn, Cu}}

\author{P. J. Baker}
\email{peter.baker@stfc.ac.uk}
\affiliation{Clarendon Laboratory, University of Oxford, Parks Road, Oxford. OX1 3PU. United Kingdom}
\affiliation{ISIS Facility, STFC Rutherford Appleton Laboratory, Chilton, Oxon. OX11 0QX. United Kingdom}
\author{T.~Lancaster}
\affiliation{Clarendon Laboratory, University of Oxford, Parks Road, Oxford. OX1 3PU. United Kingdom}
\author{I.~Franke}
\affiliation{Clarendon Laboratory, University of Oxford, Parks Road, Oxford. OX1 3PU. United Kingdom}
\author{W.~Hayes}
\affiliation{Clarendon Laboratory, University of Oxford, Parks Road, Oxford. OX1 3PU. United Kingdom}
\author{S.~J.~Blundell}
\affiliation{Clarendon Laboratory, University of Oxford, Parks Road, Oxford. OX1 3PU. United Kingdom}
\author{F.~L.~Pratt}
\affiliation{ISIS Facility, STFC Rutherford Appleton Laboratory, Chilton, Oxon. OX11 0QX. United Kingdom}
\author{P.~Jain}
\affiliation{Department of Chemistry and Biochemistry, Florida State University, Tallahassee, Florida 32306-4390, United States of America}
\affiliation{Department of Materials Science and Metallurgy, University of Cambridge, Cambridge CB2 3QZ, United Kingdom}
\author{Z.-M.~Wang}
\affiliation{College of Chemistry and Molecular Engineering, Peking University, Beijing 100871, China.}
\author{M.~Kurmoo}
\affiliation{Institut de Physique et Chimie des Matériaux de Strasbourg, CNRS-UMR 7504,
23 Rue du Loess, BP 43, F-67034 Strasbourg Cedex 2, France.}
\date{\today}

\begin{abstract}
Muon spin relaxation measurements are reported on samples of dimethylammonium metal formates containing magnetic divalent nickel, cobalt, manganese, and copper ions. These hybrid organic-inorganic perovskites exhibit weak ferromagnetism and are, apart from the copper system, multiferroics with well separated magnetic and antiferroelectric transitions. We use muons to follow the sublattice magnetization, observing the effect of the spin reorientation transitions below $T_{\rm N}$ and the criticality approaching $T_{\rm N}$. The multiferroic samples have three-dimensional antiferromagnetic interactions, but the copper sample shows quasi-one-dimensional behavior due to its Jahn-Teller distorted structure, with a ratio of its inter- and intrachain exchange constants $\vert J^{\prime}/J \vert \simeq 0.037$.
\end{abstract}


\maketitle

Designing materials to achieve functional goals is one of the major challenges of modern condensed matter physics and materials chemistry.~\cite{rao08}
This is achieved either fortuitously or by careful consideration of how chemical substitutions tune the properties of materials. While the chemical variety possible when combining elements is considerable, combining them with organic groups leads to far greater variety and control of the interactions within the materials. Such hybrid organic-inorganic materials (also called metal-organic frameworks or coordination polymers) offer the desired ability to design and synthesize functional compounds, particularly with respect to gas storage and catalysis.~\cite{yaghi03,cheetham06,cheetham07,kurmoo09} 

In magnetic systems, organic groups can be used to vary the dimensionality and strength of the interactions between transition metal ions.~\cite{manson09} Work has also led to non-centrosymmetric structures for both optical~\cite{holman01,evans02} and ferroelectric~\cite{jain08} applications. Combining these magnetic and structural properties can lead to multiferroicity, as has recently been demonstrated in \chem{[(CH_{3})_{2}NH_{2}]{\it M}(HCOO)_{3}} with {\it M}~$=$~Ni, Mn, Co, and Fe, where the antiferroelectric ordering occurs in the range 160-185~K and the magnetic ordering in the range 8-36~K.~\cite{jain09,zwang,wang,sanchez10} These dimethylammonium metal formates adopt the \chem{ABX_3} perovskite structure with {\it M}$^{2+}$ ions (B) bridged by \chem{(HCOO)^-} ions (X), and \chem{[(CH_{3})_{2}NH_{2}]^-} ions (A) at the center of a \chem{ReO_{3}} type cavity.~\cite{jain09}
The antiferroelectric phase transition appears to be driven by the ordering of hydrogen bonds linking the dimethylammonium cations.~\cite{jain09,sanchez10} This mechanism also occurs in the prototypical ferroelectric potassium dihydrogen phosphate (KDP)~\cite{slater41} and can be contrasted with the lone pair displacement evident in \chem{BiFeO_3} and the magnetically driven ferroelectricity in compounds such as \chem{TbMn_{2}O_5}.~\cite{fiebig05} The antiferroelectric and magnetic transitions are well separated, suggesting weak coupling between these effects and placing these compounds in the category of Type-I multiferroics like \chem{BiFeO_3} and \chem{YMnO_3}.~\cite{khomskii09}

\begin{figure}[t]
\includegraphics[width=\columnwidth]{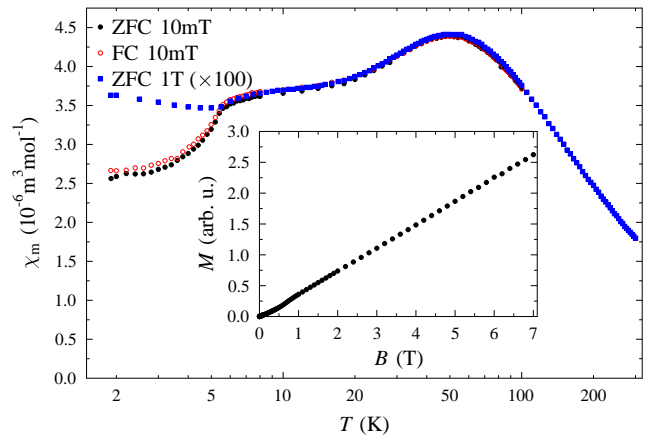}
\caption{\label{structure} (Color online)
Magnetic susceptibility of \chem{[(CH_{3})_{2}NH_{2}]Cu(HCOO)_3} at fields of $10$~mT and $1$~T. (Inset) Magnetization at $1.9$~K as a function of field.
}
\end{figure}

The copper analogue of these compounds has been known for a considerable time,~\cite{sletten} but its magnetic properties have not previously been studied in detail. Unlike the other members of this series it undergoes a Jahn-Teller distortion above room temperature that orders the central amine moiety. Bulk measurements shown in Fig.~\ref{structure} suggest antiferromagnetic order below $T_{\rm N} = 5.5$~K, a critical field for removing the three-dimensional magnetic order of $H_{c} = 0.66$~T, and a Bonner-Fisher-like susceptibility~\cite{bonner64} appropriate to a quasi-one-dimensional magnet above $20$~K, with $g=2.280(3)$ and an intrachain exchange constant $J = 81(1)$~K. (Our $J$ value is equivalent to $2J/k_{\rm B}$ in Ref.~\onlinecite{bonner64}.)

Here we report muon spin relaxation ($\mu$SR) experiments carried out on samples of \chem{[(CH_{3})_{2}NH_{2}]{\it M}(HCOO)_{3}} (with {\it M}~$=$~Ni, Co, Mn, and Cu). These provide microscopic information on the sublattice magnetization, the spin reorientation transitions, and the critical behavior. We also compare our results to previous magnetic susceptibility measurements~\cite{zwang,wang} and examine the effect of the Jahn-Teller transition on the magnetic properties of the copper sample.

Samples of \chem{[(CH_{3})_{2}NH_{2}]{\it M}(HCOO)_{3}}, where {\it M} = Ni, Co and Mn, were synthesized under solvothermal conditions at $140^{\circ}$~C. Metal chloride salts (5~m~mol) were dissolved in a $60$~mL solution of $50$~vol~\% dimethylformamide (DMF) in water, and the solution was transferred into a Teflon-lined autoclave. This was heated for 3~days at $140^{\circ}$~C. The autoclaves were air cooled, and the supernatants were transferred into a glass beaker for room temperature crystallization. After a further 3 days, transparent cubic crystals were obtained. Samples of \chem{[(CH_{3})_{2}NH_{2}]Cu(HCOO)_{3}} were synthesized by mixing hydrated copper chloride, formic acid, and dimethylamine in DMF at room temperature resulting in blue cubic crystals. 

The $\mu$SR experiments~\cite{blundell99} were carried out at the Paul Scherrer Institute using the GPS spectrometer and at the ISIS Pulsed Muon Facility using the EMu spectrometer. 
The muon spin polarization is followed as a function of time by measuring the asymmetry in the count rate of decay positrons, $A(t)$, in two detectors on opposite sides of the sample.
Commensurate magnetic order generally leads to oscillations in the muon decay asymmetry as muons precess around the local fields at their stopping site(s) with the damping of these oscillations determined by the distribution of these local fields and their fluctuations. Muons that stop with their spin along the direction of the local field will not precess, but can have their spins flipped by fluctuating magnetic fields.

\begin{figure}[t]
\epsfig{file={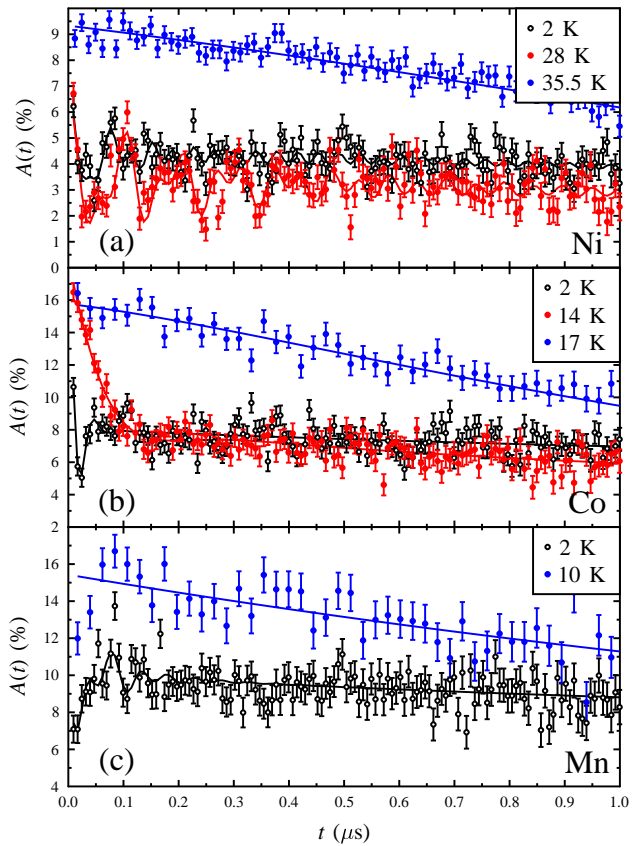},width={\columnwidth}}
\caption{\label{raw} (Color online) 
Muon decay asymmetry data for \chem{[(CH_{3})_{2}NH_{2}]{\it M}(HCOO)_{3}}: (a) $M=$Ni, (b) $M=$Co, and (c) $M=$Mn. 
Fewer muons are counted for data sets recorded above the magnetic ordering transitions and the graphs correct for this by using larger histogram bins. The Ni data have a lower asymmetry because the initial muon spin polarization was aligned differently with the detectors. Fits to the data are described in the text.
}
\end{figure}

\begin{figure*}
\epsfig{file={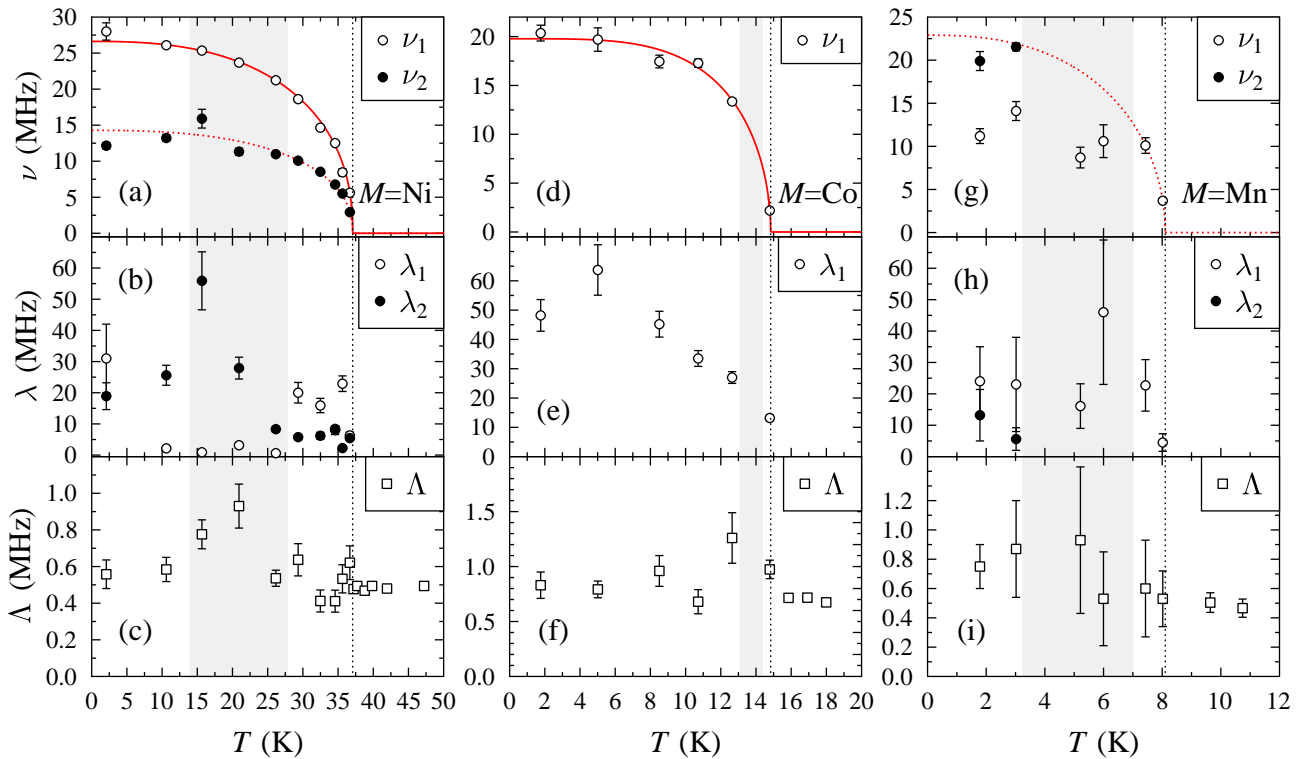},width={2\columnwidth}}
\caption{\label{parameters} (Color online)
Parameters derived from fitting Eq.~\ref{fitfunc} to the $\mu$SR data for \chem{[(CH_{3})_{2}NH_{2}]{\it M}(HCOO)_{3}} shown in Fig.~\ref{raw}. Panels (a), (d), and (g) show the oscillation frequencies $\nu_i$. Fitted solid lines in (a) and (d) correspond to Eq.~\ref{nuT} with the parameters listed in the text and Table~\ref{table}. The dashed line in (g) is a guide to the eye with $\alpha =2.5$ and $\beta =0.5$, highlighting the distinct behavior for {\it M} = Mn. (b), (e), and (h) show the linewidths $\lambda_i$ for each oscillating component. (c), (f), and (i) show the exponential relaxation rate $\Lambda$. The N\'{e}el temperatures are denoted by the vertical dashed lines and the shaded areas show the regions over which the spin-reorientation transitions occur.
}
\end{figure*}

\begin{figure}[t]
\epsfig{file={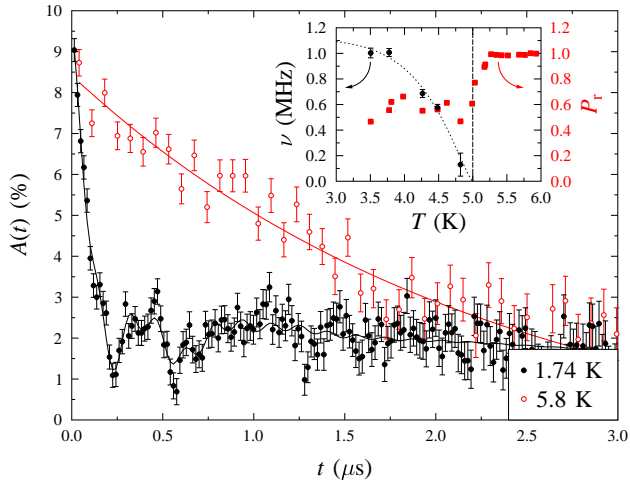},width={\columnwidth}}
\caption{\label{cu} (Color online)
Raw data recorded on Cu-formate at PSI with the inset showing the primary oscillation frequency and the fractional amplitude of the relaxing component $P_{\rm r} = A_{\parallel}/A_{\rm total}$ observed in the ISIS data. 
}
\end{figure}

The data shown in Figs~\ref{raw} and~\ref{cu} were analysed using the WiMDA program.~\cite{wimda} It was found that similar fitting functions were suitable for describing the data on each of the four samples, based on the general form:
\begin{equation}
A(t) =
\sum^{N}_{i} A_{i} e^{-\lambda_{i}t}\cos(2\pi\nu_{i}t)
+ A_{\parallel} e^{-\Lambda t} + A_{\rm bg}.
\label{fitfunc}
\end{equation}
The first $N$ terms describe $N$ damped oscillations. The $A_{\parallel}$ term describes the exponential relaxation for muon spins with their direction along that of the local field at their stopping site, which are depolarized by spin fluctuations. The final term describes the temperature-independent contribution to the asymmetry from muons stopping outside the sample. The observed ratio $(\sum^{N}_{i} A_{i})/A_{\parallel} \simeq 2$ indicates that all the samples are magnetically ordered throughout their volume. Above the magnetic ordering transition there is no oscillatory signal and, as is generally the case in paramagnets, the data are well described by an exponential relaxation, with rate $\Lambda$. The parameters resulting from these fits are shown in Figs~\ref{parameters} and~\ref{cu}.

For well-defined oscillation frequencies that varied continuously below $T_{\rm N}$ we fitted the temperature dependence to the phenomenological function: 
\begin{equation}
\nu_i (T) = \nu_i (0) [1-(T/T_{\rm N})^{\alpha}]^{\beta},
\label{nuT}
\end{equation}
where $\alpha$ describes the $T \rightarrow 0$ trend and $\beta$ describes the trend approaching $T_{\rm N}$. 


Ni-formate provided the most clearly resolved muon data of the three samples [Fig.~\ref{raw}~(a)] with two oscillating components of similar amplitude, suggesting that the muons are stopping at two structurally similar but magnetically inequivalent sites. However, below about $25$~K the oscillation frequencies cease to be in proportion which, together with step changes in the linewidths around this temperature, signals that the spin-reorientation transition begins well above the temperature of $15$~K found in ac susceptibility measurements.~\cite{wang} This suggests that the change in the canting at the reorientation transition involves a range of fluctuation timescales that appear within the time-windows of the two probes at different temperatures, with the shorter timescale of $\mu$SR being influenced at higher temperature. Fitting $\nu_1(T)$ with Eq.~\ref{nuT} gives the parameters $\nu_{1}(0) = 26.6(1)$~MHz, $T_{\rm N} = 37.07(5)$~K, $\alpha = 2.6(1)$, and $\beta=0.43(2)$. This is plotted in Fig.~\ref{parameters}~(a) together with a rescaled dotted line assuming $\nu_{2}(0) = 14.3$~MHz for comparison with the $\nu_2$ values measured.


The Co-formate data showed only one oscillating component below $T_{\rm N}$, with a similar amplitude to the total oscillating amplitude in the other samples. Even though the linewidth $\lambda_1$ is greater than the oscillation frequency $\nu_1$, the damped oscillation is clear at all temperatures below $T_{\rm N}$, as can be seen in Fig.~\ref{raw}~(b). Fitting $\nu_1 (T)$ to Eq.~\ref{nuT} gave: $\nu_1 (0) = 19.8(7)$~MHz, $T_{\rm N} = 14.84(6)$~K, $\alpha = 4.0 \pm 1.5$, and $\beta = 0.51(12)$. 
Our value of $T_{\rm N}$ is in excellent agreement with the bulk measurements, but we do not see a clear step in $\nu_1$ analogous to the step found in $M(T)$ at $13.1$~K.~\cite{wang} We do see a slight increase in the relaxation rate $\Lambda$ in the appropriate temperature region and the temperature dependence of the oscillation frequency is unusually flat below the spin reorientation. The linewidth $\lambda_1$ seems to roughly follow $\nu_1$ meaning that the local field distribution is dominating the linewidth, rather than spin fluctuations. This suggests a large number of almost magnetically equivalent muon stopping sites, which is fully consistent with the weak ferromagnetism observed in bulk measurements.


Below 4~K, the Mn-formate data shown in Fig.~\ref{raw}~(c) display two oscillating components with $A(\nu_1) \gg A(\nu_2)$. Above 5~K only one oscillation frequency, $\nu_1$, is observed, with a large linewidth giving strongly overdamped oscillations. This change is consistent with the kinks in the ZFC susceptibility.~\cite{wang} Fitting with Eq.~\ref{nuT} does not give statistically significant parameters. Wang {\em et al.} found $T_{\rm N} = 8.5$~K,~\cite{wang} which is consistent with the $\mu$SR data. We see that the longitudinal relaxation rate $\Lambda$ is close to being temperature-independent. The linewidth $\lambda_1$ behaves similarly to $\nu_1$, again suggesting a large number of nearly magnetically equivalent muon stopping sites, although this near equivalence is disrupted by cooling through the spin-reorientation transition.

Our two sets of measurements on Cu-formate are shown in Fig.~\ref{cu}. We investigated the low-temperature behaviour at PSI, where the higher time resolution allows both oscillation frequencies present to be resolved, and the behaviour around $T_{\rm N}$ in more detail at ISIS. Perhaps the most obvious difference between these data and those in the other compounds is the lower oscillation frequencies, due to the smaller magnetic moment of the copper ions. In the base temperature data we are able to resolve three oscillation frequencies of $6.44(6)$, $2.34(6)$ and $0.85(4)$~MHz. The lower two frequency components each contribute around $45$~\% of the oscillating amplitude. It is not possible to resolve all three frequencies in the higher temperature data, but the oscillations clearly persist to $T_{\rm N} \sim 5$~K, consistent with the bulk data (see Fig.~\ref{structure}). The different pattern of frequencies is likely to be due to the Jahn-Teller distorted structure rather than any change in the magnetic structure. The ISIS data taken around $T_{\rm N}$ show a sharp drop in the initial asymmetry at $5$~K as the highest oscillation frequency leaves the ISIS time window, giving a sharper measure of the transition temperature, and we see the lowest oscillation frequency, which remains within the ISIS time window, tending smoothly to zero at the transition.

\begin{table}[t]
\caption{\label{table}
Properties of \chem{[(CH_{3})_{2}NH_{2}]{\it M}(HCOO)_{3}}. The spins, $S$, derived from the magnetic susceptibility data agree with the values expected from the formal oxidation states. $J$ values are estimated using Rushbrooke and Wood's model,~\cite{randw} except for Cu-formate where the primary exchange constant, $J$, is taken from fitting the susceptibility data to the Bonner-Fisher form~\cite{bonner64} and the interchain coupling, $J^{\prime}$, is estimated using an empirical model.~\cite{yasuda05} 
The oscillation frequencies, $\nu$, are taken from the largest amplitude component in each sample, using fits to Eq.~\ref{nuT} for Ni and Co, and the $1.8$~K data for Mn and Cu.
}
\begin{ruledtabular}
\begin{tabular}{lcccc}
{\it M} & Mn & Co & Ni & Cu \\
\hline
$S$ & 5/2 & 3/2 & 1 & 1/2 \\
$T_{\rm N}$~(K) & $8.5$ & $14.84(6)$ & $37.07(5)$ & $5.5$ \\
$J$, $J^{\prime}$~(K) & -0.34, -- & -1.38, -- & -6.44, -- & -81, -3 \\
$\nu$~(MHz) & 12 & $19.8(7)$ & $26.62(9)$ & 2.34 \\
\end{tabular}
\end{ruledtabular}
\end{table}

We summarize our results in Table~\ref{table}. Looking for trends in the behavior of the three multiferroic samples, we see that $T_{\rm N}$ decreases as the moment size grows. This indicates that the details of the indirect magnetic exchange through the ligands have a greater influence on the ordering temperature than the moment size. A similar trend has been found in related compounds.~\cite{kelihu} The internal fields do not appear to follow the moment size either, most likely due to the slightly different moment directions relative to the muon sites. Our estimates for the magnetic exchange constants on the basis of the magnetic ordering temperatures~\cite{randw} are in reasonable agreement with those in Ref.~\onlinecite{wang} (on the basis of the Curie-Weiss constant), except for Co-formate where spin-orbit coupling is probably affecting the high-temperature behavior. The three multiferroic samples are therefore well described by these three-dimensional models.

In conclusion, the $\mu$SR data on these samples allow us to track the sublattice magnetization, rather than the bulk magnetization due to the weak ferromagnetism that is seen in magnetic susceptibility measurements.~\cite{wang}  The temperature dependence of the muon oscillation frequencies shows that the critical behavior in the Ni- and Co-formate samples is consistent with a mean-field description. The spin-reorientations influence the $\mu$SR data at higher temperature than in ac susceptibility measurements,~\cite{wang} strongly suggesting they are continuous and associated with a broad range of timescales. In contrast to the three-dimensional and multiferroic members of this series, we find that Cu-formate is quasi-one-dimensional, and the ratio of its inter- and intrachain exchange constants derived from the ordering temperature and the high-temperature susceptibility is $\vert J^{\prime}/J \vert \simeq 0.037$. This is very similar to the ratio in \chem{KCuF_3} ($\vert J^{\prime}/J \vert \simeq 0.052$), where the behavior is also determined by a Jahn-Teller distortion.~\cite{satija80} 
Parts of this work were performed at the Swiss Muon Source, Paul Scherrer Institute, Villigen, CH and the ISIS Facility, UK. We thank A. Amato for experimental assistance; A. K. Cheetham, H. W. Kroto, and N. Dalal for useful discussions, and the EPSRC and STFC (UK) for financial support.


\end{document}